\documentclass[aps,twocolumn,aps,showpacs]{revtex4}

\usepackage[dvips]{graphicx}

\begin{document}

\title{Spectral Dependence of Coherent Backscattering of Light in a
Narrow-Resonance Atomic System}
\author{D.V. Kupriyanov, I.M. Sokolov, and N.V. Larionov}
\affiliation{Department of Theoretical Physics, State Technical University, 195251,
St.-Petersburg, Russia}
\author{P. Kulatunga, C.I. Sukenik, S. Balik, and M.D. Havey}
\affiliation{Department of \ Physics, Old Dominion University, Norfolk, VA 23529}
\date{\today }

\begin{abstract}
We report a combined theoretical and experimental study of the spectral and
polarization dependence of near resonant radiation coherently backscattered
from an ultracold gas of $^{85}$Rb atoms. Measurements in a $\pm 6$ $MHz$
range about the $5s^{2}S_{1/2}\rightarrow 5p^{2}P_{3/2}$ $F=3\rightarrow
F^{\prime }=4$ hyperfine transition are compared with simulations based on a
realistic model of the experimental atomic density distribution. \ In the
simulations, the influence of heating of the atoms in the vapor,
magnetization of the vapor, finite spectral bandwidth, and other nonresonant
hyperfine transitions are considered. \ Good agreement is found between the
simulations and measurements. \
\end{abstract}

\pacs{32.80.-t, 32.80.Pj, 34.80.Qb, 42.50.-p, 42.50.Gy, 42.50.Nn}
\maketitle


\section{Introduction}

Coherent wave scattering effects in disordered media display an
extraordinary variety of phenomena which are of both fundamental
and practical concern. \ Of particular interest is that coherent
wave scattering shows a broad universality which makes possible a
qualitatively similar description for different types of wave
excitation in a variety of media. These range, as an illustration,
from enhancement of light scattering off the lunar regolith and
the rings of Saturn on the one hand \cite{Mish}, to explanation of
peculiarities in propagation of waves in the solid earth on the
other \cite{POAN}. In addition, coherent wave scattering is a
useful technique for diagnosing the average properties of
scatterers in turbid media, and for assessing relatively thin
surface layers in biological and mechanical materials
\cite{Sheng,LagTig,POAN}. The propagation of light waves in
natural photonic materials such as opal gives the valuable
semiprecious gemstone its highly valued beauty. Of fundamental
scientific
importance, coherent wave scattering was first recognized by Anderson \cite%
{Anderson} in the context of interference of electron wave scattering in
conductors. As the scattering mean free path decreases and becomes shorter
than a characteristic length on the order of the wavelength, wave diffusion
slows as a result of wave interference. The limiting case where diffusion
ceases is called strong localization, where the propagating wave becomes
spatially localized inside the medium. For electromagnetic radiation \cite%
{Sheng,LagTig}, two recent reports of strong localization have been made,
one in the optical regime \cite{Wiersma1}, and the other for microwave
radiation \cite{Chabanov1}. A major long-term and fundamental goal of the
research presented here, and of other researchers in the field, is to attain
strong localization of light, but in an ultracold atomic vapor.

Quite recently, coherent multiple light scattering has been observed in
ultracold atomic gases, which form a unique and flexible medium for
fundamental studies and practical applications \cite%
{Labeyrie1,Labeyrie2,Kulatunga1,Bidel}. In all cases, the essential physical
mechanisms are due to interferences in multiple wave scattering from the
components of the medium; under certain not very stringent conditions the
interferences survive configuration averaging, thus generating macroscopic
observables. First observations and initial explanations for electromagnetic
radiation were of the so-called coherent backscattering (CBS) cone in
disordered media \cite{Ishimaru,Wolf, Albeda}. For radiation incident on a
diffusive medium, the effect manifests itself as a spatially narrow ($\sim $%
1 mrad) cusp-shaped intensity enhancement in the nearly backwards direction %
\cite{Sheng,LagTig}. As electromagnetic waves are not scalar, the detailed
shape and size of the enhancement depends on the polarization of the
incident and the detected light. Nevertheless, for classical radiation
scattering from a $^{1}S_{0}\rightarrow $ $^{1}P_{1}$ atomic transition, the
largest possible interferometric enhancement is to increase the intensity by
a factor of two.

Atomic gases, because they have exceptionally high-Q resonances, and because
the light scattering properties may be readily modified by light
polarization or intensity, atomic density, and applied external fields,
represent an interesting and flexible medium in which to study the role of
multiple scattering. However, to achieve the full potential of atomic
scatterers as a practical medium for such studies, it is necessary to
significantly cool the atoms, in order to suppress the dephasing effects of
atomic motion. Coherent backscattering interference has, in fact, been
measured in $^{85}$Rb \cite{Labeyrie1,Kulatunga1} and Sr \cite{Bidel}, and
quite successfully modeled for resonant and near-resonant scattering as well %
\cite{CBSth1,KCBSth1,KCBSth2,Jonckheere}. Measurements have also been made
of the magnetic field dependence of the coherent backscattering line shape %
\cite{Labeyrie3}, and of the time-dependence, for a particular geometry, of
light scattered in the coherent scattering regime \cite{Kaiser}. However,
there remains a significant range of physical parameters associated with the
various processes which have not yet been fully explored. \ Among these are
the influence of light intensity, nonzero ground state multipoles such as
alignment or orientation, cooperative multi-atom scattering associated with
higher atomic density, and more general geometries for time-dependent
studies. \ In the present report we concentrate attention on another
variable, that being the dependence of the coherent backscattering
enhancement on detuning of the probe beam from exact resonance. \ It is
clear that non-resonant excitation of the atomic sample results in a smaller
optical depth (and associated larger transport mean free path) of the medium %
\cite{CBSth1,KCBSth1}. \ However, theoretical and experimental results
presented here reveal that other more subtle effects, including
far-off-resonance optical transitions, heating of the vapor by multiple
light scattering and self-magnetization of the vapor during the CBS phase,
can have significant effects on the spectral variation of the CBS
enhancement. \

In the following sections we first present an overview of the physical
system, including how atomic samples are prepared and characterized and a
brief review of measurements of coherent backscattering from an atomic
vapor. \ This is followed by a summary of the approach to simulate coherent
multiple scattering in an ultracold atomic gas. \ We then present our
experimental and theoretical results, with focus on various mechanisms that
can influence the spectral variation of the coherent backscattering
enhancement factor.

\section{Overview of Physical System}

\subsection{Preparation and description of ultracold atomic sample}

Preparation of the ultracold atomic $^{85}$Rb sample used in the
measurements described in this paper has been described in detail elsewhere %
\cite{Kulatunga1}, but for completeness will be briefly reviewed here. \ The
samples are formed in a vapor-loaded magneto-optical trap (MOT) which is
operated in a standard six-beam configuration. The trapping laser is detuned
a frequency of -2.7$\gamma $ from resonance, where $\gamma \sim 5.9$ $MHz$,
is the natural linewidth of the $F=3\rightarrow F^{\prime }=4$ hyperfine
transition in $^{85}$Rb. Laser light for the MOT is derived from an
injection locked diode laser (Sanyo DL7140-201) which is slaved to a master
laser (Hitachi HG7851G). The master laser is locked to a crossover peak
produced in a Doppler-free saturated absorption spectrometer. Laser locking
is achieved by dithering the master laser current and demodulating the
saturation absorption spectrum with a lock-in amplifier. In order to produce
the required light for hyperfine repumping, the slave laser is microwave
modulated to produce a sideband at the wavelength corresponding to the $%
F=2\rightarrow F^{\prime }=3$ hyperfine transition. Light exiting the slave
passes through an acousto-optic modulator (AOM), which is used as an optical
switch, and subsequently coupled into a single mode fiber optic patchcord.
The combination of the AOM switching and fiber coupling results in an $\sim $%
65 dB attenuation of the trapping laser light. After exiting the fiber, the
trapping light is split into three beams and sent to the MOT. Each beam
contains $\sim $3.3 mW of light and is retroreflected, generating an average
$\sim $19 mW in the center of the chamber.

In order to ascertain the number and density of confined atoms, we employ
absorption and fluorescence imaging. We find that the MOT is not completely
spherical \cite{Kulatunga1,CBSth1}, but rather is somewhat `cigar-shaped'
having $1/e^{2}$ Gaussian radii of 1.1 mm and 1.38 mm, where the radius is
defined according to the density distribution $n(r)=n_{0}\exp
(-r^{2}/2r_{0}^{2})$, $n_{0}$ being the peak density. This distribution
results in an optical depth through the center of the MOT of about 6, where
the optical depth $b$ is defined as resulting in an attenuation of the
incident intensity by a factor $e^{-b}$. We determine the peak optical depth
by direct measurement of the transmitted CBS light intensity through the
central region of the MOT. In these measurements, probing of the density
takes place when the MOT lasers are off, for they result in a significant
excited state fraction, decreasing the measured optical depth. For a
Gaussian atom distribution in the MOT, the optical depth is given by $b=$ $%
\sqrt{2\pi }\sigma _{0}n_{0}r_{0}$, where $\sigma _{0}$ is the cross-section
for light scattering\ \cite{Metcalf}. With the values given above and an
average Gaussian radius $r_{0}=1.2$ $mm$, we calculate that the MOT contains
approximately $4.3\times 10^{8}$ atoms and has a peak density $n_{0}=$ $%
1.6\times 10^{10}$ atoms-cm$^{-3}$. \ Note that these parameters are large
enough to insure an optical depth large enough for coherent multiple
scattering, but that the density is not so large as to necessitate
consideration of cooperative pair scattering in the vapor. \

\begin{figure}
  \includegraphics[width=3.0 in]{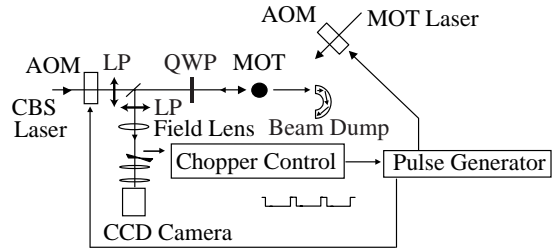}\\
  \caption{\label{Figure1}Schematic diagram of the coherent
backscattering apparatus. Shown in the figure is an acousto optic
modulator (AOM), magneto optic trap (MOT), linear polarizer (LP),
quarter wave plate (QWP), and a charge coupled device (CCD)
camera.}
\end{figure}

The vapor-loaded MOT is formed in a custom-made stainless steel
ultrahigh vacuum (UHV) chamber that is pumped by both an ion and
titanium sublimation pump. The UHV chamber is fitted with a
stainless-steel sidearm containing a valvable and heated Rb
reservoir. \ Because we are observing light which is backscattered
from our sample, it is critical that all other backscattered
reflections are suppressed. A major source of unwanted
back-scattered light is from the vacuum viewports on the MOT
chamber. In order to minimize this light, we installed wedged
optical quality windows having a ``V''-type antireflection (AR)\
coating at 780 nm on the probe-laser (described in the following
section) entrance and exit ports. The AR coating results in less
than 0.25\% reflectivity at 780 nm. Further, the window through
which the probe laser beam enters is mounted on a UHV bellows,
allowing us to better direct unwanted reflections from entering
the charge-coupled device (CCD) detector. We also found it
necessary to replace the standard window on the CCD camera with a
wedged and near-infrared AR coated window in order to suppress
interference fringe formation in the CCD images. \

\subsection{Measurement of atomic coherent backscattering}

We present in this section a brief overview of the coherent
backscattering apparatus used to obtain the experimental results
reported here. Further details may be found in Kulatunga,
\textit{et al. }\cite{Kulatunga1}, where the experimental
apparatus used in experiments to study coherent radiative transfer
in an ultracold gas of $^{85}$Rb \cite{Kulatunga1} is described. A
schematic diagram of the arrangement is shown in Figure 1. \ There
the external light source used in the experiment is provided by an
external cavity diode laser that is stabilized by saturated
absorption to a crossover resonance associated with hyperfine
components of the 5s $^{2}$S$_{1/2}$ $ \rightarrow $ 5p
$^{2}$P$_{3/2}$ transition. With reference to Figure 2, which
shows relevant hyperfine transitions in $^{85}$Rb, the laser may
be tuned several hundred MHz from nearly any hyperfine resonance
in $^{85}$Rb by a standard offset locking technique using an
acousto-optic modulator. Detuning from resonance is defined by
$\Delta =\omega _{L}-\omega _{0}$,
where $\omega _{L}$ is the CBS laser frequency and $\omega _{0}$ is the $%
F=3\rightarrow F^{\prime }=4$ resonance frequency. The laser bandwidth is a
few hundred kHz, and the typical output power is $\sim $5 mW. The laser
output is launched into a single mode polarization preserving fiber and then
beam expanded and collimated by a beam expander to a $1/e^{2}$ diameter of
about 8 mm. The polarization of the resulting beam is selected and then the
beam passed through a nonpolarizing and wedged 50-50 beam splitter that
passes approximately half of the laser power to the atomic sample. The
backscattered radiation is directed by the same beam splitter to a field
lens of 45 cm focal length, which condenses the light on the focal plane of
a liquid nitrogen cooled CCD camera. The diffraction limited spatial
resolution is about 100 $\mu rad$, while the polarization analyzing power is
greater than 2000 at 780 nm. There are four polarization channels that are
customarily studied in coherent backscattering. For linearly polarized input
radiation, two of these correspond to measuring the backscattered light in
two mutually orthogonal output channels. This is readily achieved by
removing the quarter wave plate, as shown in Figure 1, and rotating the
linear polarization analyzer located before the field lens. For input
radiation of definite helicity, that being generated by the linearly
polarized input and the quarter wave plate, the other two channels
correspond to the helicity of the backscattered radiation. \ This is
similarly measured by rotation of the linear polarizer just before the field
lens. The instrumentation as described, with some modifications to suppress
the intense trapping beam fluorescence, has been previously used to study
coherent backscattering in ultracold atomic gases and in solid and liquid
samples as well \cite{Kulatunga1}.

\begin{figure}
 \includegraphics[width=3.0 in] {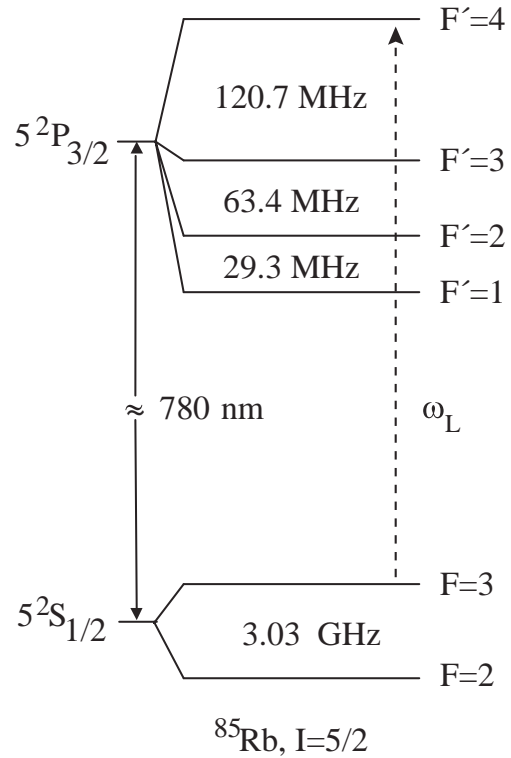}\\
 \caption{\label{Figure2}Hyperfine energy levels of relevant
transitions in atomic $^{85}$Rb.}
\end{figure}

Measurements of the backscattered light is made by exposing the
ultracold atoms to the CBS laser light for an interval of 0.25 ms
temporally centered in a 5 ms dark interval during which the MOT
lasers are turned off. \ The MOT lasers are then turned back on
for 20 ms, which is sufficiently long that the cold atom sample is
reconstituted. \ This procedure is repeated for 300 s, which
constitutes a single experimental run. \ A run of 300 s with the
MOT absent allows measurement of the background, which is
principally due to hot atom fluorescence excited by the CBS laser.
\ Attenuation of the CBS laser by the MOT during the data taking
phase, which reduces the amount of background during the
backscattering run, in comparison with the background phase, is
accounted for by auxiliary measurements of the MOT attenuation of
the CBS laser intensity. \ Finally, the saturation parameter for
the CBS laser is less than s = 0.08 on resonance, which with the
0.25 ms measurement interval is sufficient to minimize mechanical
action of the CBS laser beam on the atomic sample. \

\subsection{Brief overview of the theoretical treatment}

A general theory of the coherent backscattering process in an ultracold
atomic gas has been developed recently by several groups \cite%
{CBSth1,KCBSth1,KCBSth2}. The theoretical development essentially maintains
the earlier conceptions of weak localization in the atomic scattering
problem \cite{Shlyap}, and takes into account the influence of the optical
depth and sample size on the character of the coherent backscattering cone.
In spite of the fact that the basic ladder and interference terms,
describing the process, have a similar structure in all the theoretical
approaches, there are certain types of accompanying physical phenomena which
can become more important as more detailed experimental or theoretical
spectral analysis is considered.

In our earlier theoretical approach \cite{CBSth1}, the general analytical
development was realized by a Monte-Carlo simulation of coherent multiple
scattering in an ultracold ($T<50$ $\mu K$) gas of $^{85}$Rb atoms confined
in a magneto optical trap. \ The simulation was closely matched to the
experimental density distribution and temperature conditions as described in
the previous paragraphs. \ The radiation field frequency was selected to be
in the vicinity of the $F=3\rightarrow F^{\prime }=4$ hyperfine transition,
and to have polarization states and a weak-field intensity corresponding to
the experimental realization. The effects of sample size, and the spatial
and polarization dependence of the coherent backscattering cone were
considered in detail. \ Some aspects of the spectral variation of the
coherent backscattering enhancement factor were also considered, including
the surprisingly-strong influence of the far-off-resonance $F=3\rightarrow
F^{\prime }=3$ and $F=3\rightarrow F^{\prime }=2$ hyperfine transitions.
However, other physical effects can have a profound influence on the
spectral variations, and we consider some of those in the present report. \
Among these, for currently achievable laboratory conditions, are (i) heating
of the atomic gas by multiple scattering of the probing light source, (ii)
optical pumping effects initiated by the probe or MOT lasers, and (iii) the
influence of the finite bandwidth of the probe laser. Each of these effects
has been quantitatively ignored in earlier studies, since their role is not
so crucial to calculations of the basic characteristics of the CBS process.
Of particular interest is the influence of atomic motion and internal
polarization variables on the spectral variations of the CBS enhancement. \
We point out that to properly account for these factors, it is necessary to
consider the influence of the mean field on both attenuation and dispersion
of the multiply scattered light, and to include also the anisotropic Green's
function for light propagating along a chain of scatterers. \ As is seen in
the following section, inclusion of some such effects, in isolation or
combination, may well be essential to better agreement between experimental
and theoretical results.

Finally we emphasize that the simulations are made for conditions quite
close to those in the experiment. These conditions include sample size,
temperature, shape and density, and the characteristic intensity of the CBS
laser beam. \ These conditions are such that cooperative scattering may be
neglected, and such that saturation of the atomic transition is also
negligible. In simulations of thermal effects, and of the influence of
atomic magnetization on the coherent backscattering enhancement, more severe
conditions are used in order to illustrate possible range of influence of
the effects. \

\section{Experimental and Theoretical Results}

In this section we present experimental and theoretical results associated
with backscattering of near-resonance radiation from ultracold atomic $^{85}$%
Rb. \ First we present experimental measurements of the spectral variation
of the coherent backscattering enhancement, in a range of approximately $\pm
6MHz$, as a function of detuning from the $F=3\rightarrow F^{\prime }=4$
hyperfine transition. \ These results are directly compared to theoretical
simulations, made with inclusion of the influence of off-resonant hyperfine
transitions and considering an ultracold sample not at absolute zero. \
Second, we present simulations of several effects which should generally be
considered when modelling coherent backscattering from ultracold atomic
vapors.

\subsection{Spectral variation of the CBS enhancement: experimental results}

\begin{figure}
  \includegraphics{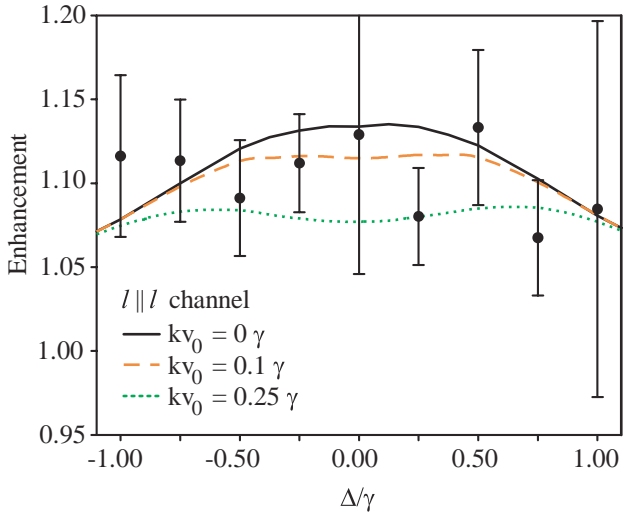}\\
  \caption{\label{Figure3}Comparison of experimental and
theoretical enhancement spectra in the \emph{l} $\|$ \emph{l}
polarization channel. \ Theoretical spectra show modification by
Doppler broadening, which is varied from $kv_{0}=0$ to
$kv_{0}=0.25\protect\gamma $,
in an ensemble of ${}^{85}$Rb atoms having a peak density of $%
n_{0}=1.6\times 10^{10}\;cm^{-3}$ and a Gaussian radius
$r_{0}=1\;mm $.}
\end{figure}

\begin{figure}
  \includegraphics{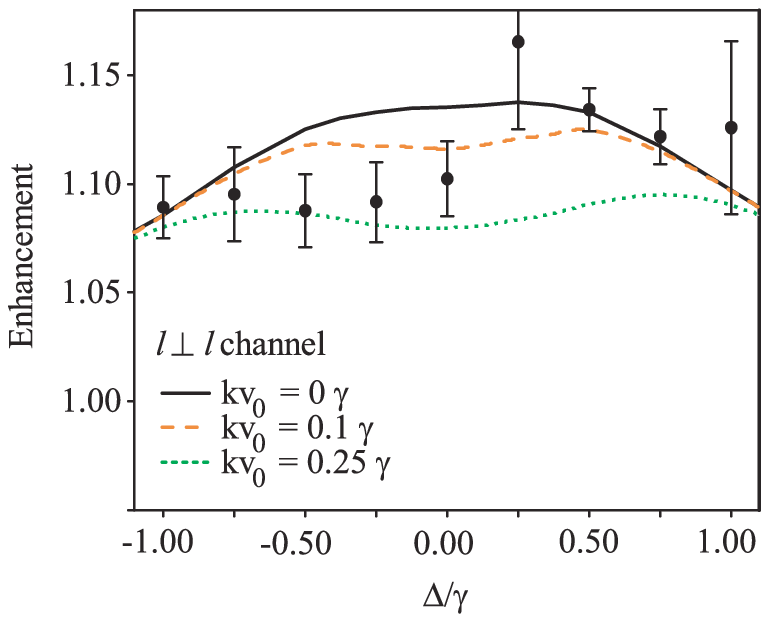}\\
  \caption{\label{Figure4}Comparison of experimental and
theoretical enhancement spectra in the \emph{l} $\perp $ \emph{l}
polarization channel. \ Theoretical spectra show modification by
Doppler broadening, which is varied from $kv_{0}=0$ to
$kv_{0}=0.25\protect\gamma $,
in an ensemble of ${}^{85}$Rb atoms having a peak density of $%
n_{0}=1.6\times 10^{10}\;cm^{-3}$ and a Gaussian radius
$r_{0}=1\;mm $.}
\end{figure}

\begin{figure}
  \includegraphics{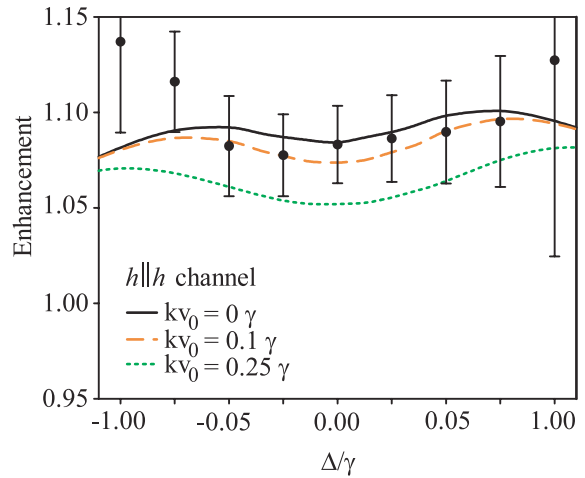}\\
  \caption{\label{Figure5}Comparison of experimental and
theoretical enhancement spectra in the helicity preserving
($h||h)$ polarization channel. \ Theoretical spectra show
modification by Doppler broadening, which is varied from
$kv_{0}=0$ to $kv_{0}=0.25\protect\gamma $,
in an ensemble of ${}^{85}$Rb atoms having a peak density of $%
n_{0}=1.6\times 10^{10}\;cm^{-3}$ and a Gaussian radius
$r_{0}=1\;mm$.}
\end{figure}

Measurements of the variation of the coherent backscattering
enhancement with detuning of the CBS laser, in a $\pm 6MHz$ range
around the $F=3\rightarrow F^{\prime }=4$ hyperfine transition,
are shown in Figures 3-6. \ The measurements have a typical
uncertainty on the order of 2\%, this being due to a combination
of statistical uncertainty due to counting statistics in the
spatial intensity measurements, but also an estimated
uncertainty due to the cone fitting procedure, as described previously \cite%
{Kulatunga1}. \ In addition, there is residual noise in the spatial
distribution of backscattered light due to speckle in the $l$ $||$ $l$ and $%
h\perp h$ channels; slight variations in speckle appearing in
background scattered light from run to run does not completely
average to zero. \ This effect is responsible for the somewhat
larger fluctuations in the extracted enhancement factors for these
two polarization channels. \ There is a small systematic reduction
of the peak enhancement due to the finite spatial resolution of
the backscattering polarimeter, an effect which arises from
smoothing of the nearly cusped shaped CBS cone near its peak by
the finite spatial resolution of the instrument. \ This is
accounted for by using a Lorentzian model of the spatial response
and the CBS cone, which allows an estimate of the amount of
reduction. \ Justification for this is made by the fact that the
spatial variation of the simulated cones is to a good
approximation described by a Lorentzian, as is the measured
spatial response of the experimental apparatus. \ In our case,
accounting for this effect amounts to a maximum decrease in the
peak enhancement $\sim 0.01$ in the enhancement for the narrowest
cones, which appear in the $h$ $||$ $h$ data. This estimated
correction is not made to the data in Figs. 3-6. On the scale of
the figures, there is negligible uncertainty in the detuning
measurements.

\begin{figure}
  \includegraphics [width = 3.0 in] {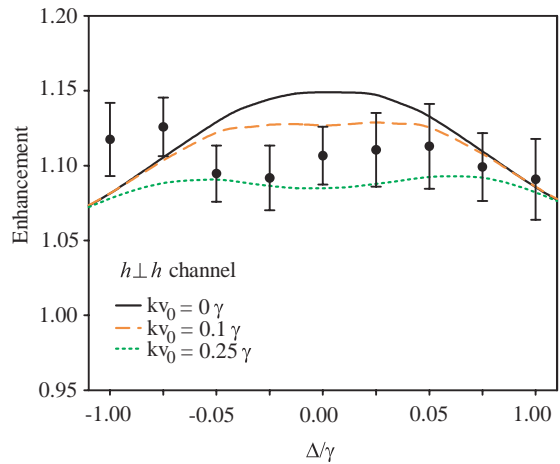}\\
  \caption{\label{Figure6}Comparison of experimental and
theoretical enhancement spectra in the helicity non-preserving
$(h\perp h)$ polarization channel. \ Theoretical spectra show
modification by Doppler broadening, which is varied from
$kv_{0}=0$ to $kv_{0}=0.25\protect\gamma $,
in an ensemble of ${}^{85}$Rb atoms having a peak density of $%
n_{0}=1.6\times 10^{10}\;cm^{-3}$ and a Gaussian radius
$r_{0}=1\;mm$.}
\end{figure}

Also shown in the figures are simulations of the enhancement for
several different values of the average Doppler shift of the
atoms, measured in units of the natural spectral width $\gamma $.
\ It appears, from comparison of the experimental data and the
simulations that the data is better described by inclusion of some
nonzero average heating of the vapor \ on the order of a few
hundred kHz. \ Note that in a following section we model the
influence of the finite bandwidth of the CBS laser; it is seen
there that the decrease in enhancement on resonance, as seen in
Figures 3-6, cannot be explained by that mechanism.

\subsection{Influence of dynamical heating}

The initial temperature of the atomic ensemble is $\sim $ $50\;\mu K$, which
makes negligible any possible spectral manifestations caused by atomic
motion. However, during the interaction time, the probe light produces a
certain mechanical action on the atoms. The radiation force associated with
the probe light can accelerate the atoms and heat them to temperatures where
the Doppler broadening and shift can become comparable with the natural
linewidth. It is important to recognize that the initial scattering event
transfers momentum from the CBS laser to the atomic ensemble, but that
subsequent scattering of the light deep within the sample is more nearly
isotropic, resulting in some effective heating of the atoms during the CBS
data taking phase. \ Although the dynamical process is complex, and is
currently under study, we present here a short discussion of this process by
comparing several scanning spectra averaged over an equilibrium Maxwell
distribution of atom velocities, that is for different temperatures and
respectively for different Doppler widths. The drift velocity of the atomic
cloud, which also exists but was ignored in our calculations, leads (to a
good approximation) to a Doppler shift of all the spectral dependencies into
the blue wing with respect to the laser frequency $\omega _{L}$.

In addition to the experimental data, there are also shown in Figures 3-6
calculated spectral variations of the enhancement factor for all four
polarization channels. In the graphs, the velocity is indicated as a
fraction of the natural width of the atomic transition, $kv_{0}$ ($v_{0}=%
\sqrt{2k_{B}T/m}$ is the the most probable velocity in the atomic ensemble).
\ It is seen from these graphs that the shape of the spectra becomes
significantly modified, even for an average Doppler breadth of $0.1\gamma $,
where $\gamma $ is the natural atomic width. \ Similar results are also
obtained in the linear polarization channels. \ The overall trend suggests
that dynamical heating, or some other mechanism which modifies the spectrum
in a similar way, will be required to describe the experimental results. \
Of particular interest is the helicity preserving channel (Figure 5). \
Unique to this case is an increase in the enhancement, even for no atomic
motion, for moderate detunings away from exact resonance. As described in a
previous report \cite{CBSth1}, this is due to the suppressed role of
Raman-type single scattering. The asymmetry is due to the nonnegligible
influence of far off resonance hyperfine transitions on the coherent
backscattering enhancement. \ This effect is suggested in the overall
spectral trend of the data, although more precise measurements are clearly
in order. \ In the following section, we see that a much larger enhancement
increase at greater detunings can also arise from dynamical (or static)
magnetization of the vapor along the direction of propagation of the CBS
laser beam. \

\begin{figure}
  \includegraphics{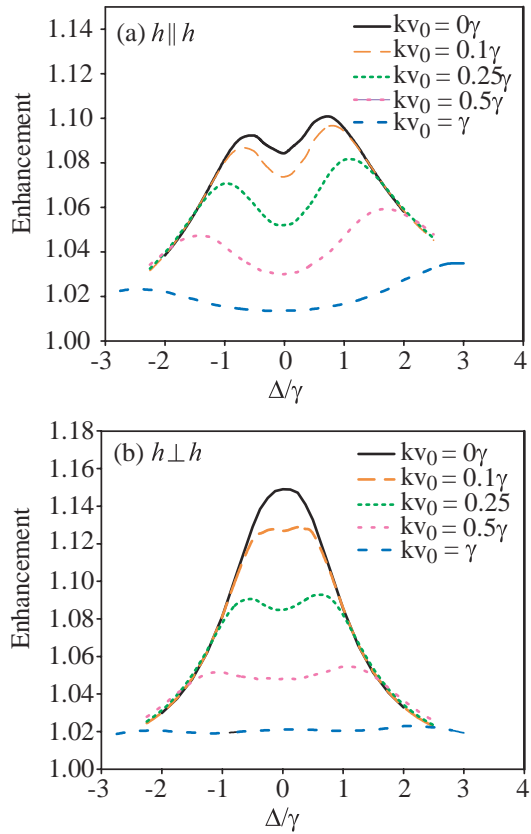}\\
  \caption{\label{Figure7}Scanning spectra of CBS
enhancement for (a) \textit{h} $||$ \textit{h} and (b) \textit{h}
$\perp $ \textit{h} polarization channels. Spectra are shown for
an average Doppler broadening
varying from $kv_{0}=0$ to $kv_{0}=\protect\gamma $, in the ensemble of $%
{}^{85}$Rb atoms with $n_{0}=1.6\times 10^{10}\;cm^{-3}$ and
$r_{0}=1\;mm$.}
\end{figure}

We finish this section by presenting for completeness theoretical results
over a wider spectral range for the detuning dependence of the CBS
enhancement. \ These are shown in Figure 7 (a) for the helicity preserving
channel and in Figure 7 (b) for the helicity nonpreserving channel. \ The
results show that there is some persistence of the CBS enhancement, even for
an average Doppler broadening on the order of the natural line width of the
atomic transition, and that this enhancement increases at larger detunings,
before falling off at the largest offsets, when single scattering becomes
dominant. \

\subsection{Optical pumping effects}

Effects on coherent radiative transport in an atomic vapor will generally
depend on the polarization of the incident light and on the nonzero ground
state multipoles in the atomic vapor. \ In our experimental arrangement, the
atoms are confined to a magneto optic trap, in which there exist generally
spatially varying hyperfine multipoles. \ However, the MOT lasers are
typically turned off for several ms before taking data in a coherent
backscattering experiment, and residual macroscopic atomic polarization
should be largely dissipated on that time scale. \ However, there can be
hyperfine multipoles generated by the CBS laser itself, dynamically
polarizing the vapor. The main argument, why there is no optical pumping
manifestation in the CBS process, comes from the reasonable assumption that
under not atypical experimental conditions the probe radiation is weak and
characterized by a small saturation parameter. However, it is quite clear
that in an ensemble of cold atoms the relaxation mechanisms in the ground
state, which mainly are collisional, play a reduced role and become even
negligible. Then, after each cycle of interaction with the polarized CBS
light, the atomic ensemble can accumulate a certain degree of polarization,
which may be either of an orientation or alignment type. Of particular
interest to us here is when the incident radiation has definite helicity,
which can magnetize the vapor along the CBS propagation direction. \ It is
quite difficult to estimate precisely the actual dynamical spatial
distribution, within the atomic cloud, of polarization generated during the
whole interaction cycle. Therefore in this section we only qualitatively
illustrate how the optical pumping effects can change a basic characteristic
of the CBS process such as the enhancement factor.

\begin{figure}
  \includegraphics[width=2.5in] {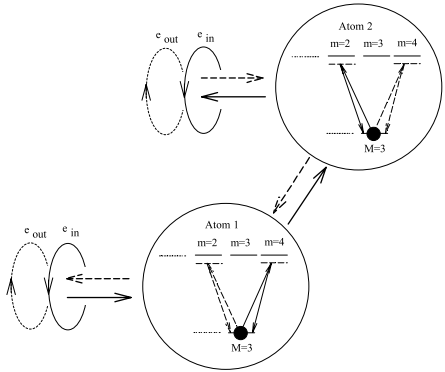}\\
  \caption{\label{Figure8}Diagram explaining the CBS
phenomenon for double scattering in an ensemble of oriented atoms
of ${}^{85}$Rb atoms. There is only one transition amplitude in
the helicity preserving channel, which leads to maximal
enhancement of backward scattered light. The direct and reciprocal
transitions and photon paths are shown by solid and dashed arrows
respectively.}
\end{figure}

Consider probing the atomic sample with positive helicity circular
polarized radiation. Let us consider further that, due to optical
pumping, the atoms become oriented only along the propagation
direction of the light beam. In steady state, following a
sufficiently long pumping time, if there is no relaxation in the
ground state, the atoms should be concentrated in the Zeeman
sublevel $F=3,M=3$, see Figures 2 and 8. Of course, this is an
idealized case which can never be precisely attained in reality,
but such a model situation is convenient for illustrative
purposes. The spectral variations of the enhancement factor for
such an oriented ensemble are shown in Figure 9 for the case of
monochromatic probe radiation and for a Gaussian-type cloud
with the peak density $n_{0}=1.6\times 10^{10}\;cm^{-3}$ and with a radius $%
r_{0}=1\;mm$.

The spectral variation of the enhancement in the helicity
preserving channel shows quite unusual behavior, in that there is
no reduction of the CBS enhancement in the spectral wings. On the
contrary, the enhancement factor approaches its maximal possible
value of two. The limiting factor of two is normally associated
with Rayleigh-type scattering on classical objects. But here we
deal with Rayleigh-type scattering under approximately attainable,
but not typical quantum conditions. This result may be explained
by a simple but fundamental property that in such a coherent
atomic ensemble there is no single-atom scattering of Raman-type\
radiation in the backward direction, which potentially could also
be a source of backscattered light in the helicity-preserving
channel. Moreover, the partial contribution of only double
scattering on oriented atoms in an optically thin sample causes
the enhancement factor to take the maximum possible numerical
value of two. This can be understood by turning to Figure 8, where
it is shown that there is only one channel, or one product of the
transition matrix elements, contributing in the scattering
amplitude of the double Rayleigh scattering, that is in the
helicity preserving channel. These are ideal conditions to observe
maximal enhancement in the CBS process. In higher orders there are
several partial contributions and not all of them can interfere. \
This, as usual, leads to essential reduction of the interference
contribution to the total intensity of scattered light. Due to
such reduction the enhancement factor considerably decreases in
the spectral domain near the resonance scattering, as shown in
Figure 9. Thus in the wings of the helicity preserving curve in
the graph of Figure 9 a unique situation is revealed when
\textit{in an optically thin medium under special conditions the
enhancement factor can increase to its maximal value}.

\begin{figure}
  \includegraphics[width=2.5in]{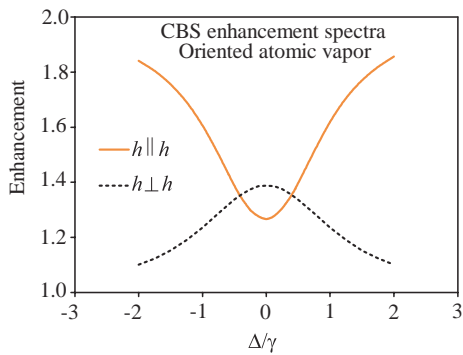}\\
  \caption{\label{Figure9}Scanning spectra of enhancement
for circular polarized probe light in an ensemble of atoms of
${}^{85}$Rb atoms with 100\% orientation in the direction the
light propagation. The spectra were calculated for a Gaussian type
atomic cloud with $n_{0}=1.6\times 10^{10}\;cm^{-3}$ and
$r_{0}=1\;mm$.}
\end{figure}

\begin{figure}
  \includegraphics[width=2.5in]{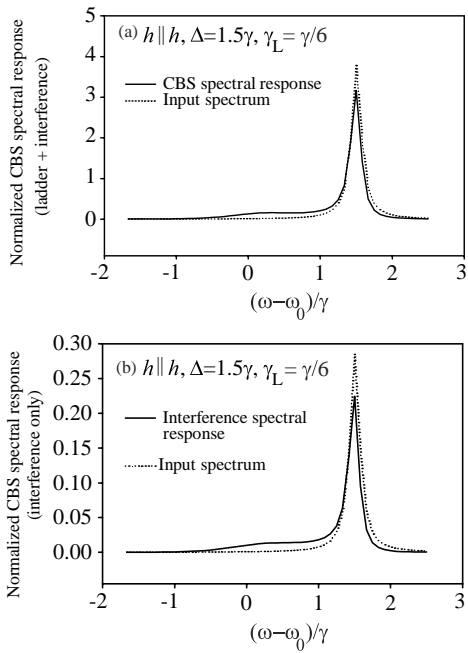}\\
  \caption{\label{Figure10}The output spectral response of
the CBS light when the input circular polarized laser radiation,
modeled by a Lorentzian spectrum (\ref{t1}) with $\protect\omega _{L}=\protect\omega %
_{0}+1.5\protect\gamma $ and $\protect\gamma _{L}=\protect\gamma
/6$, is tuned in the blue wing of the $F=3\rightarrow F=4$ optical transition of $%
{}^{85}$Rb. The first graph (a) shows the distortion of the input
Lorentzian profile (dotted curve) for the total ladder and
interference contribution; it is normalized to the total output
intensity of the CBS light. The second graph (b) shows the
distortion for the interference term only; it is normalized
according to the corresponding enhancement factor. Both the
dependencies relate to the same helicity preserving channel.}
\end{figure}

The multiple scattering in the helicity non-preserving channel
shows more ordinary behavior. The disappearance of CBS in the
wings is caused by the dominating contribution of single
scattering events as far as the sample becomes optically thin. We
point out that there is, for this polarization channel also, a
certain increase of the maximal value of enhancement in comparing
with a non-oriented atomic ensemble. Here, as in the linear
polarization channels, we see that optical pumping phenomenon
leads to some quantitative but not qualitative changes in
observation of the CBS process. However, the combined results of
the numerical simulations presented in this and the previous
section suggest that the experimental results may be essentially
modified by the combined influence of thermal and optical pumping
effects.

\subsection{The finite bandwidth of the probe light spectrum}

In an experiment, the CBS probe laser operates ideally in a single-mode
regime and its spectral bandwidth is much less than the natural relaxation
rate of the atoms. But in reality the difference is not necessarily so great
to completely ignore the spectral distribution of the laser radiation.
Typically in our experiments on spectral scanning the sample consisting of
rubidium atoms, which have a resonance line natural decay rate $\gamma \sim
5.9\;MHz$, the laser radiation has normally a bandwidth of less than $1\;MHz$%
. For the multiple scattering process in higher orders, the scanned spectral
profile of the sample is formed as a successive overlap of individual
profiles per scattering event, the effective output shape reveals a much
narrower spectral variance than $\gamma $. Thus the bandwidth of the laser
mode can become comparable with the spectral inhomogeneity in the sample
spectrum associated with partial contributions of the higher scattering
orders.

This can be quantitatively discussed with the following model of a
quasi-monochromatic single mode laser radiation. To define the basic
parameters we approximated the assumed homogeneously broadened spectrum of
the CBS laser by a Lorentzian profile
\begin{equation}
I(\omega )=I\,\frac{\gamma _{L}}{\left( \omega -\omega _{L}\right)
^{2}+\left( \gamma _{L}/2\right) ^{2}}  \label{t1}
\end{equation}%
where $\omega _{L}$ and $\gamma _{L}$ are the carrier frequency and the
spectral bandwidth of the laser radiation respectively. $I$ is the total
intensity of the incident laser radiation. The spectrum obeys to the
following normalization condition
\begin{equation}
I\;=\;\int_{-\infty }^{\infty }\frac{d\omega }{2\pi }\,I(\omega )  \label{t2}
\end{equation}%
where in the quasi-monochromatic approximation there is no difference
between $0$ and $-\infty $ in the lower limit of this integral. The basic
idea in this case is that, in comparison with purely monochromatic
radiation, the spectral response of the initially symmetric but broadened
input spectral profile (\ref{t1}) should be significantly distorted by the
sample due to effects of multiple scattering.

In Figure 10 we show, in the helicity preserving channel, the
output spectral response in the backward direction when the input
circular
polarized laser radiation, modeled by the spectrum of Eq. (\ref{t1}) with $%
\omega _{L}=\omega _{0}+1.5\gamma $ and $\gamma _{L}=\gamma /6$, is tuned in
the blue wing of the $F=3\rightarrow F^{\prime }=4$ optical transition of $%
{}^{85}$Rb with resonant frequency $\omega _{0}$. The first graph plotted in
Figure 10 (a) shows the distortion of the input Lorentzian profile for the
total ladder and interference contribution, the intensity being normalized
to the total output intensity of the CBS light. In turn, the second graph
depicted in Figure 10 (b) shows the distortion of the interference term
only.\ This term is normalized according to the corresponding enhancement
factor $X_{EF}$ [to $(X_{EF}-1)/X_{EF}$]. Both the dependencies relate to
the same helicity preserving channel. It is clearly seen that the output
spectral profile becomes asymmetric because of the influence of resonance
scattering near the atomic transition in higher orders of multiple
scattering. It\ may be less obvious, but there is also a small but not
negligible difference between the two spectral dependences, which explains
why in our experiment the spectral probe of the sample with scanning carrier
frequency $\omega _{L}$ near the resonance can be sensitive to the spectral
bandwidth of the CBS laser.

\begin{figure}
  \includegraphics{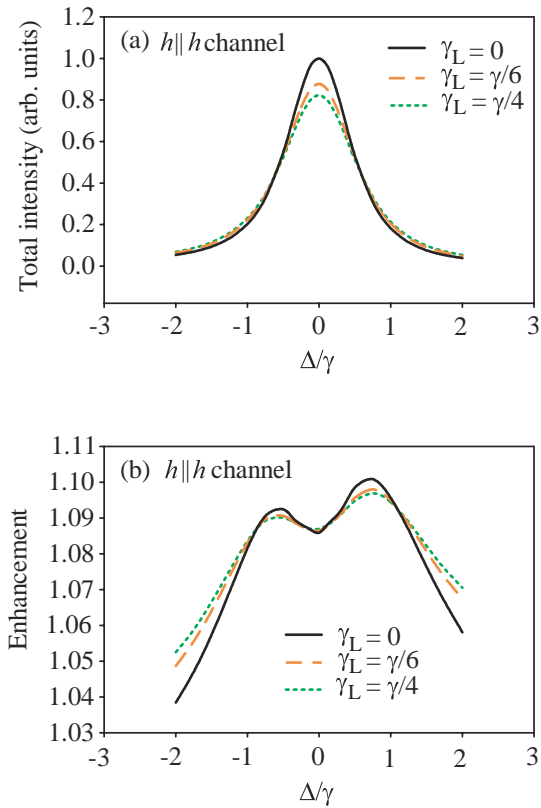}\\
  \caption{\label{Figure11}Scanning spectra of the
intensity (a) and of the enhancement factor (b) in the helicity
non-preserving channel for the quasi-monochromatic laser
radiation, with $\protect\gamma _{L}=\protect \gamma
/4,\,\protect\gamma /6,\,0$. The spectra were calculated for a
Gaussian type atomic cloud of ${}^{85}$Rb atoms with
$n_{0}=1.6\times 10^{10}\;cm^{-3}$ and $r_{0}=1\;mm$.}
\end{figure}

\begin{figure}
  \includegraphics{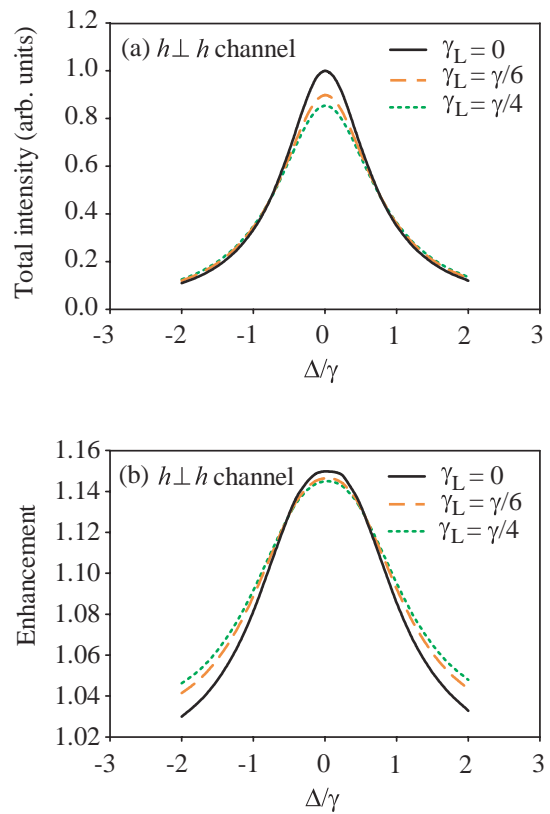}\\
  \caption{\label{Figure12}Scanning spectra of the
intensity (a) and of the enhancement factor (b) in the helicity
preserving channel for quasi-monochromatic laser radiation with
$\protect\gamma _{L}=\protect\gamma /4,\,\protect\gamma /6,\,0$.
The spectra were calculated for a Gaussian type atomic cloud of
${}^{85}$Rb atoms with $n_{0}=1.6\times 10^{10}\;cm^{-3}$ and
$r_{0}=1\;mm$.}
\end{figure}

The influence of this effect is illustrated, for the helicity preserving and
non-preserving channels, in Figures 11 and 12. \ There it is shown that the
spectra of the total intensity and of the enhancement factor, generated by
scanning of the frequency $\omega _{L}$, reveals different spectral
behavior, particularly in the wings of scanned profiles. These are
calculated results for the helicity polarization channels, but similar
behavior takes place for the linear polarization channel. At first sight
this spectral divergence appears as a rather weak effect, but we believe it
should not be ignored in precise comparison of the experimental data with
numerical simulations. Particularly, it can be important in a realistic
estimation of the background, since such a spectral washing in the probe
radiation response can be important in the interpolation procedure of the
CBS cone to its wing. Indeed, the higher orders of multiple scattering
contribute to the formation of the central portion of the CBS cone, but the
role of second order scattering is more important in its wings. As we see
for large spectral detunings, the correct estimation of the enhancement
factor in higher orders of multiple scattering is rather sensitive to the
spectral distribution of the probe radiation.

\section{Summary}

A combined theoretical and experimental study of spectral
variations in the coherent backscattering enhancement factor, for
a very narrow band resonance system, has been reported. \
Experimental data taken over a range of two atomic natural widths
about direct atomic resonance suggests spectral variations in the
peak value of the CBS enhancement. \ Simulations indicate that the
combined influence of heating of the atomic ensemble, and optical
pumping of the Zeeman sublevels in the $F=3$ ground level during
the coherent backscattering data taking phase, can qualitatively
account for the effects. \ The simulations of the CBS process
examined the influence of atomic motion, in a thermal equilibrium
model, on the spectral variation of the enhancement factor. \ A
model case of magnetization of the vapor due to optical pumping
was also considered. \ It was found that these two factors could
explain variations in the CBS enhancement observed in the
experiments. \ The simulations which considered the influence of
atomic magnetization predicted a remarkable result; the classical
CBS maximum enhancement of two could be closely approached for a
strongly magnetized atomic sample. \ Finally, it was shown, by
simulation of the influence of the spectral bandwidth of the CBS
probe laser, that even a quite small laser line width, in
comparison to the natural width of the atomic transition, can
influence significantly the CBS enhancement in the wings of the
atomic resonance line.

\begin{acknowledgments}
We acknowledge informative discussions with Robin Kaiser.\ Financial support
for this research was provided by the National Science Foundation
(NSF-PHY-0099587, NSF-INT-0233292), by the North Atlantic Treaty
Organization (PST-CLG-978468), by the Russian Foundation for Basic Research
(01-02-17059), and by INTAS (INFO 00-479). \ D.V.K. would like to
acknowledge financial support from the Delzell Foundation, Inc.
\end{acknowledgments}

\end{document}